\documentclass{emulateapj}
\usepackage{amsmath}
\usepackage{graphicx}
\usepackage{amsfonts}
\usepackage{natbib}
\usepackage{color}
\usepackage{epstopdf}
\usepackage{graphicx}
\usepackage{multirow}

\begin{document}

\title[The X-ray QPO in Mrk 766]{An X-ray periodicity of $\sim$1.8 hours in a narrow-line Seyfert 1 galaxy Mrk 766}
\author{Peng Zhang\altaffilmark{1,3}, Peng-fei Zhang\altaffilmark{1,2},  Jing-zhi Yan\altaffilmark{1}, Yi-zhong Fan\altaffilmark{1,4},  and Qing-zhong Liu\altaffilmark{1,4}}
\altaffiltext{1}{Key Laboratory of Dark Matter and Space Astronomy, Purple Mountain Observatory, Chinese Academy of Sciences, Nanjing 210008, China;
zhangpengfee@pmo.ac.cn, jzyan@pmo.ac.cn, yzfan@pmo.ac.cn,liuqz@pmo.ac.cn}
\altaffiltext{2}{Key Laboratory of Astroparticle Physics of Yunnan Province, Yunnan University, Kunming 650091, China}
\altaffiltext{3}{University of Chinese Academy of Sciences, Beijing, 100012, China}
\altaffiltext{4}{School of Astronomy and Space Science, University of Science and Technology of China, Hefei, Anhui 230026, China}

\begin{abstract}
In the narrow-line Seyfert 1 galaxy Mrk 766, a Quasi-Periodic Oscillation (QPO) signal with a period of $\sim6450$ s is detected in the \emph{XMM-Newton} data collected on 2005 May 31.
This QPO signal is highly statistical significant at the confidence level at $\sim5\sigma$ with the quality factor of $Q=f/\Delta f>13.6$.
The X-ray intensity changed by a factor of 3 with root mean square fractional variability of $14.3\%$.
Furthermore, this QPO signal presents in the data of all three EPIC detectors and two RGS cameras and its frequency follows the $f_{\rm QPO}$-$M_{\rm BH}$ relation  spanning from stellar-mass to supermassive black holes. Interestingly, a possible QPO signal with a period of $\sim4200$ s had been reported in the literature.
The frequency ratio of these two QPO signals is $\sim$ 3:2. Our result is also in support of the hypothesis that the QPO signals can be just transient.
The spectral analysis reveals that the contribution of the soft excess component below $\sim$ 1 keV is different between epochs with and without QPO,
this property as well as the former frequency-ratio are well detected in X-ray BH binaries, which may have shed some lights on the physical origin of our event.
\end{abstract}

\bigskip
\bigskip
\section{INTRODUCTION}
\label{sec:intro}
The narrow-line Seyfert 1 galaxies (NLS1s) are a subclass of active galactic nuclei (AGNs) that are powered by supermassive black holes (SMBH) accretion at the center of galaxies.
NLS1s are characterized by a narrow width of the broad Balmer emission line with FWHM (H$\beta) < 2000
{\rm~km~s^{-1}}$, along with strong optical $\rm Fe_{II}$ lines and weak forbidden lines.
Their X-ray emissions have rather rapid variability with respect to other sources.
Such variabilities are usually attributed to the dynamical processes in close vicinity of black holes (BHs) and thus
play an important role in revealing the radiation mechanism and structure of AGNs.

Quasi-periodic emissions are an interesting phenomena of some X-ray and gamma-ray emission sources.
The QPO signal has attracted wide attention. It is widely believed
to be related to the accretion in the innermost stable circular orbit (ISCO) around BH \citep{Remillard2006} and thus
carries important physical information of ISCO.
However, the QPO signal is rarely detected in AGNs, especially in NLS1s.
The first significant transient QPO has been detected in NLS1 galaxy RE J1034+396 \citep{Gierlinski2008}.
Recently two transient QPO signals with a frequency ratio of $\sim 2:1$ have been detected by \citet{Pan2016} and \citet{Zhang2017} in NLS1 galaxy 1H 0707-495.
Other possible detections of X-ray QPOs in AGNs have been reported in the literature as well, including
for example a $\sim3.8$ hr QPO in 2XMM J123103.2+110648 {\citep{Lin2013}},
a $\sim 2$ hr QPO  in MS 2254.9$-$3712 \citep{Alston2015}, and a QPO signal at $\sim 2.4\times10^{-4}$ Hz in a nearby NLS1s of Mrk 766 \citep[$z=0.0127$;][]{Boller2001}.

In this work we report the detection of a significant QPO signal at $\sim 1.55\times10^{-4}$ Hz
with confidence level of $\sim5\sigma$ on \emph{XMM-Newton} observation of 2005 May 31 with exposure time $>90$ ks.
Such a signal has a frequency about 2/3 times that of the one suggested in \citet{Boller2001}.
We also find some differences between the spectral components with QPO and without QPO signals,
similar to the behaviour detected in Black-Hole Binaries (BHBs). The QPO signal frequency and the mass of the SMBH of Mrk 766
is found to be consistent with the relation suggested in the former literature \citep{Remillard2006,Kluzniak2002,Zhou2010,Zhou2015,Pan2016}.
This work is organized as following: in Section 2 we describe the data analysis and show the main results,
and in Section 3 we have a summary with some discussions.

\section{Observations and Analysis}
\label{sec:Observations}

\subsection{Observations and Data reduction}
\label{subsec:make lc}
The European Space Agency's X-ray Multi Mirror mission (\emph{XMM-Newton}) has been launched On December 10th 1999.
It carries two sets of X-ray detectors including three European Photon Imaging Cameras \citep[EPIC; PN, MOS1 and MOS2;][]{Turner2001,Struder2001}
and two Reflection Grating Spectrometers \citep[2RGS;][]{Herder2001}.
The NLS1 Mrk 766 had been monitored 9 times for the long observation ($>30$ ks)
by \emph{XMM-Newton} from 2000 May to 2015 July in the full frame imaging mode.
We reduce the data and extract the science products using tool \emph{evselect} following the standard procedure\footnote{https://www.cosmos.esa.int/web/xmm-newton/sas-threads}
in the Science Analysis Software (SAS) with version of 16.0.0
provided in the \emph{XMM-Newton} science operations center \footnote{https://www.cosmos.esa.int/web/xmm-newton/sas-download}.
In data analysis, we select the events from a 40-arcsec-circle region of interest (ROI) centered at the position of RA=$12:18:26.48$ and DEC=$+29:48:46.15$
over energy band 0.2-10 keV. The events are excluded for the periods with high background flaring rate using tool \emph{tabgtigen}
by making a secondary Good Time Interval (GTI) file. The light-curves are generated with high quality science data
using the PATTERN $\le$ 12 for two MOS detectors and $\le$ 4 for PN detector in the tool \emph{evselect}.
The background light-curves are extracted with the events from a same diameter source-free circle ROI (without any X-ray source) in the same chip as source regions.
For this 9 observations, the pile-up effect is negligible.
The light-curves are evenly sampled with time-bin of 100 s.
Background subtraction, together with corrections for various sorts of detector inefficiencies, were performed with the SAS task \emph{epiclccorr}.
We combine light-curves with data from the three cameras (PN+MOS1+MOS2). We then also obtain the combined light-curves from the two RGSs detectors.
And the following time series analysis is based on these combined light-curves.
While for spectra analysis, the energy spectra from Mrk 766 and background are extracted with same regions applied to derive the light-curves
with the parameter of \emph{spectralbinsize}=15 in the tool \emph{evselect} for EPIC Cameras,
the corresponding response matrices are extracted simultaneously.
The detailed information for this step in provided in the SAS data analysis Threads \footnote{https://www.cosmos.esa.int/web/xmm-newton/sas-thread-pn-spectrum}.

\subsection{The combined light-curve analysis}
\label{subsec:Search qpo}
\begin{figure*}
\centering
\includegraphics[scale=0.5]{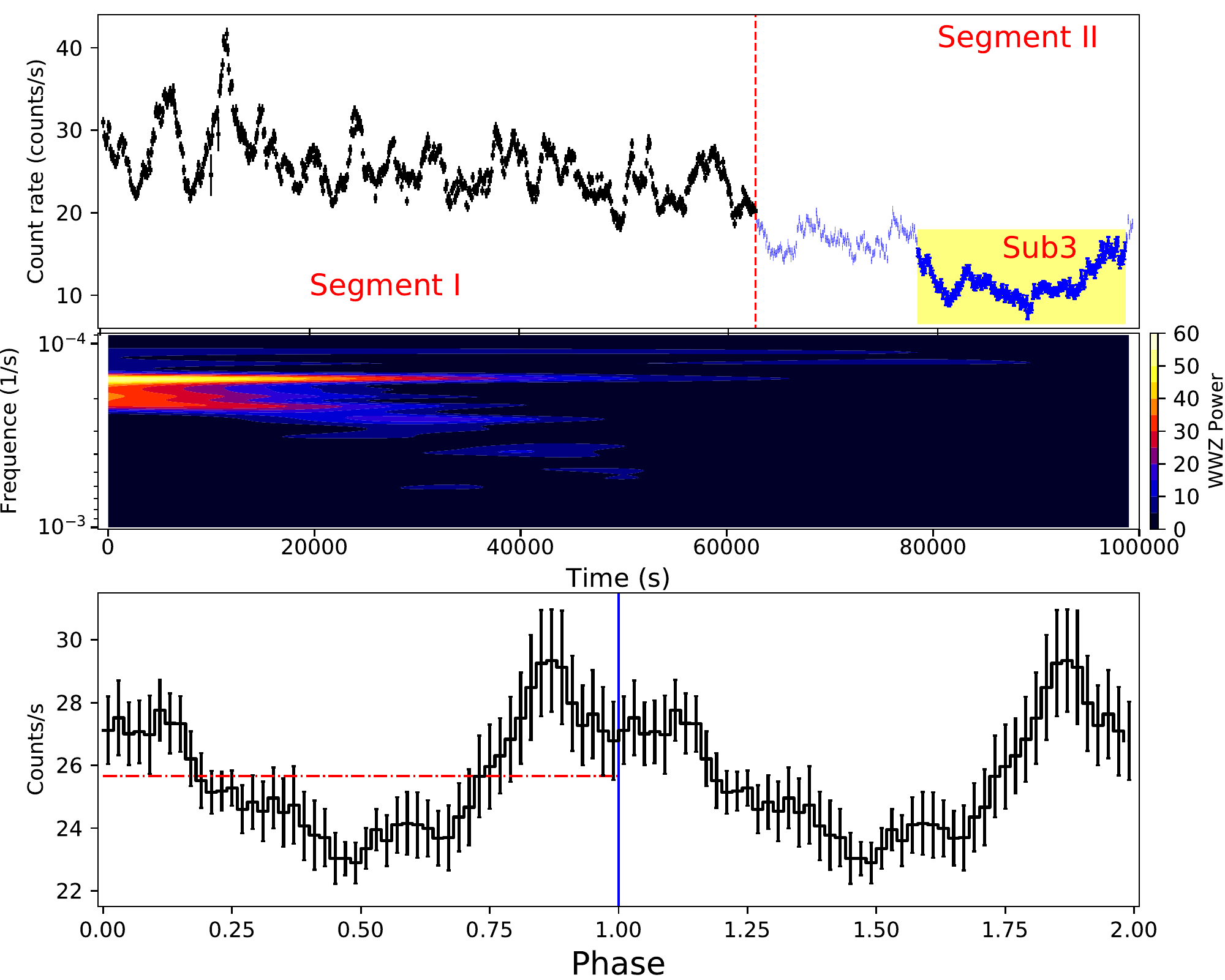}
\caption{The upper panel: \emph{XMM-Newton} EPIC light-curve of Mrk 766 in 0.2-10 keV with 100-s per bin. The light-curve is separated into two segments by a red dashed line.
              The middle panel: 2D plane contour plot of WWZ power of the whole light-curve.
              The lower panel: the pulse shape of light-curve is folded with Segment I by using the period cycle of 6451.6 s (two cycles are shown).}
\label{lc}
\end{figure*}

\begin{figure*}
\centering
\includegraphics[scale=0.39]{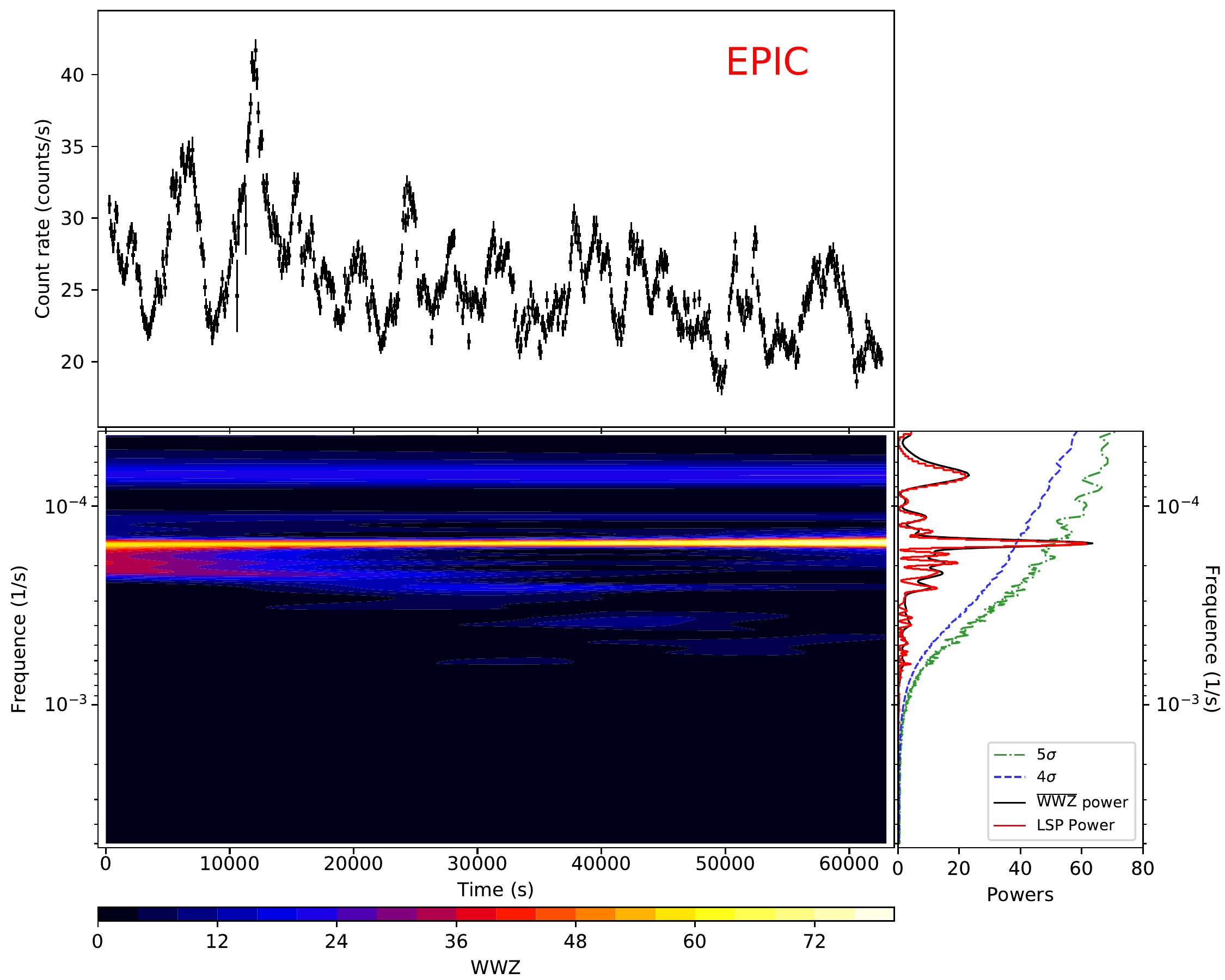}
\includegraphics[scale=0.39]{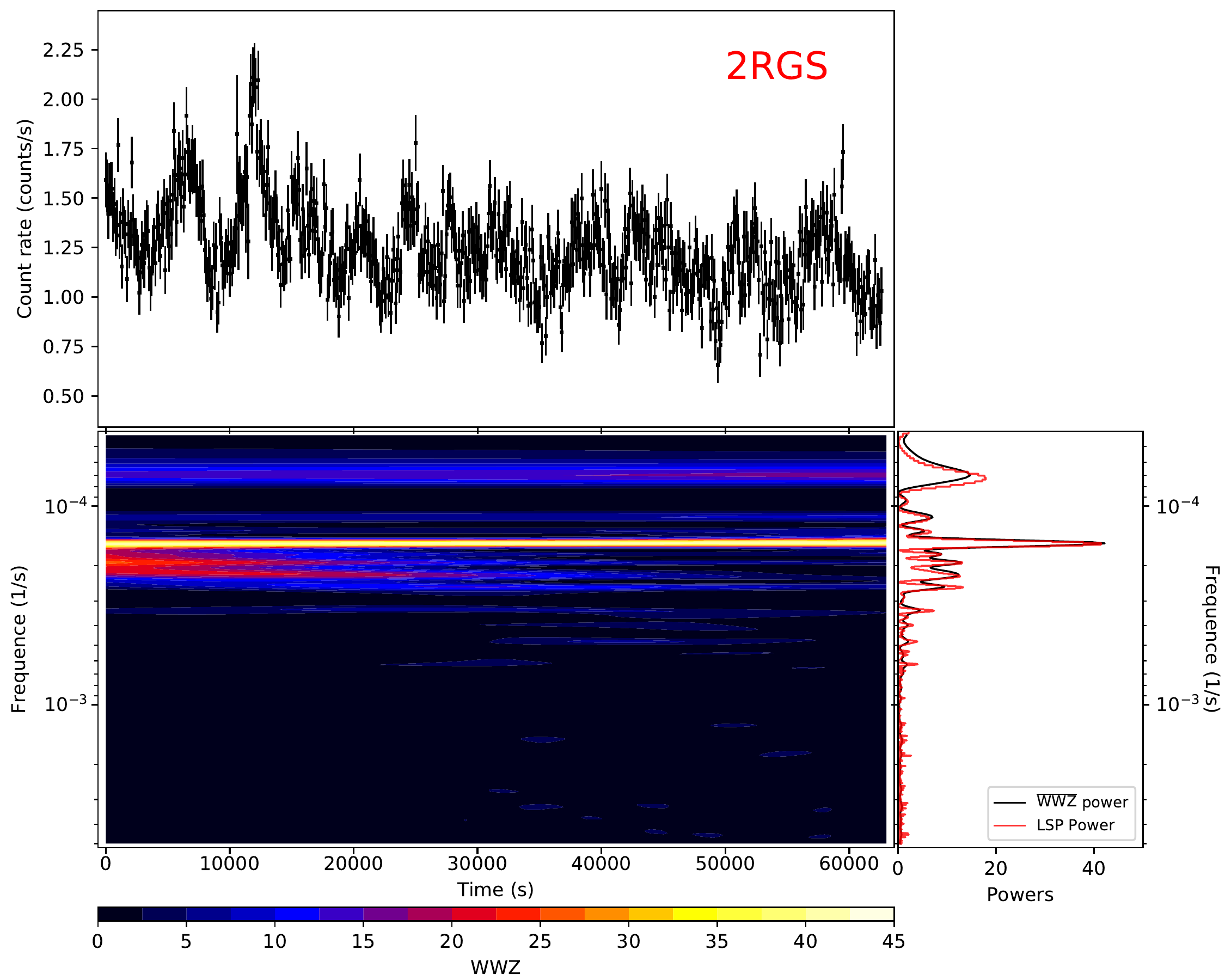}
\caption{The left image: the EPIC light-curve of Segment I is shown in the upper panel.
                                         The 2D plane contour plot of WWZ power of upper light-curve is filled in the lower left panel and the value is scaled with color.
                                         In the lower right panel, the red and black solid lines represent LSP power and time-averaged WWZ power, respectively;
                                         the confidence-level curves of 4$\sigma$ and 5$\sigma$ are shown with blue dashed and green dashed-dotted lines, respectively.
	      The right image: the 2RGS light-curve corresponding to Segment I and its WWZ power are shown in the left panel (upper and lower, respectively),
	                                 and the time-averaged WWZ power (black line) and LSP power (red line) are shown in lower right panel.}
\label{powers}
\end{figure*}

\begin{figure*}
\centering
\includegraphics[scale=0.39]{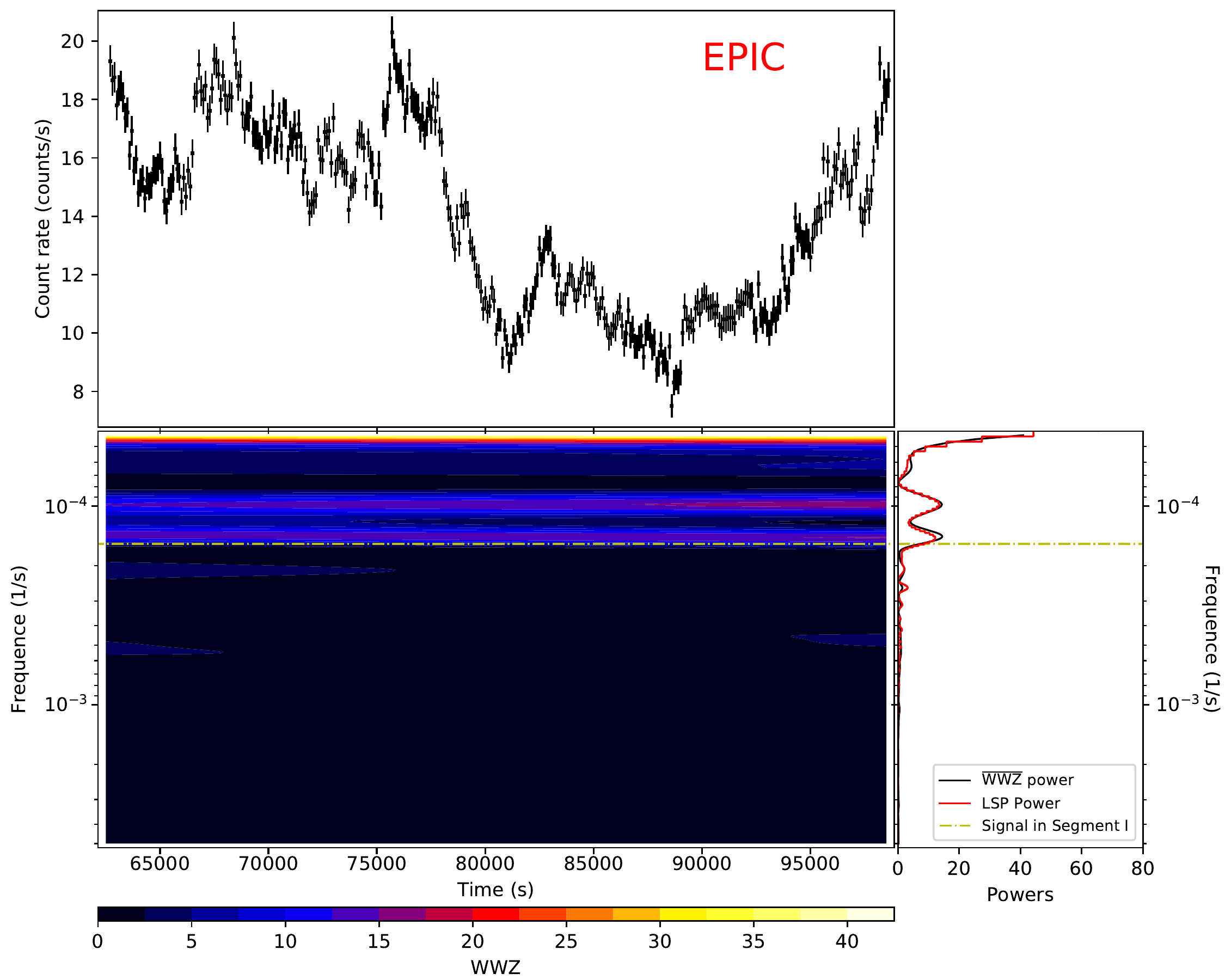}
\includegraphics[scale=0.39]{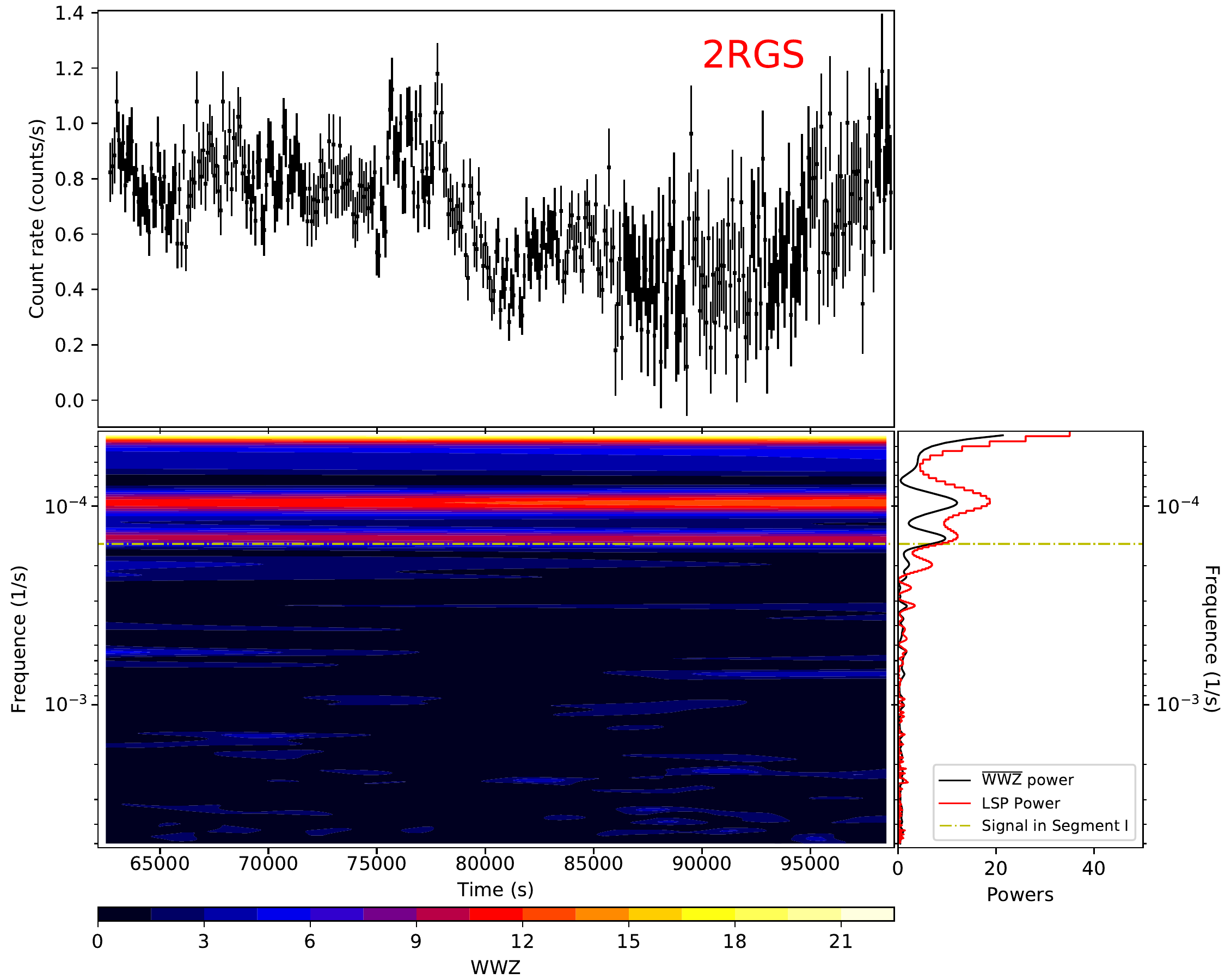}
\caption{The results of EPIC and 2RGS for Segment II. See Fig.2 for legend and the yellow dashed-dotted lines represent the signal found in the Segment I (see Fig.~\ref{powers}).}
\label{powersSII}
\end{figure*}

To search for the quasi-periodic signal, we employ two most widely used methods, the generalized Lomb-Scargle Periodogram \citep[LSP;][]{Lomb1976,Scarle1982,Zechmeister2009}
and Weighted Wavelet Z-transform \citep[WWZ;][]{Foster1996}, to obtain the power spectra of (combined EPIC and 2RGS) light-curves.
In this work the power spectra of LSP method is checked with the independent results of WWZ approach.
Particularly for the light-curves on 2005 May 31 (Obs ID: 0304030601), following the previous works  \citep{Gierlinski2008,Pan2016,Zhang2017},
we divide the EPIC light-curve into two segments (Segment I and Segment II), as shown in the upper panel of Fig.~\ref{lc}.
We focus on the power spectra of Segment I and show the results in the left image of Fig.~\ref{powers}.
In left image, the 2D plane contour plotting for WWZ power spectrum is shown in the lower left panel.
In the lower right panel, the red solid and black solid lines represent the LSP (with average Nyquist frequency $\sim$ 0.005 Hz) and time-averaged WWZ power spectra.
A strong peak at $\sim (1.55\pm0.11)\times10^{-4}$ Hz (with period cycle of 6451.6 s) is detected in both WWZ and LSP powers
(while in Segment II, the signal disappears at all, as shown in the middle panel of Fig.~\ref{lc} and in the Fig.~\ref{powersSII}).
The uncertainty of the signal is evaluated with the full width at half maximum of Gaussian-function fitting at the position of the peak.
The probability ($p_{\rm prob}$) for obtaining a power equal to or larger than the threshold from the chance fluctuation (the noise)
is $\ll 1\times10^{-15}$ \citep{HorneBaliunas1986}.
Then the probability is corrected based on the number of independent frequencies sampled (the number of trials).
The frequency resolution ($\delta f$) is $\sim 1/T_{\rm exposure}$, the frequency range ($\triangle f$) is $f_{\rm max}-f_{\rm min}$ \citep{Zechmeister2009},
and the $N$ is approximately $\sim \frac{\triangle f}{\delta f}=312$.
And the false-alarm probability ($FAP =1-[1 - p_{\rm prob}]^{N}$) is $\ll 3.2\times10^{-13}$.
To estimate the confidence level more robustly, we generate $10^{6}$ artificial light-curves based on the power spectral density (PSD)
and the probability density function of the variation of EPIC light-curve. The simulated light-curves have full properties of statistical and variability of EPIC light-curve.
To determine the best-fitting PSD, we use a bending power-law plus a constant function to model the PSD of EPIC light-curve
using a $\chi^2$ minimization technique of Minuit and get a $\chi^{2}/{\rm d.o.f}=0.3$ (where d.o.f represents the degree of the freedom).
And the function is $P(f) = Af^{-1}[1+(f/f_{bend})^{\alpha-1}]^{-1}+C$ \citep{Gonzalez-Martin2012}, where the $A$, $\alpha$, $f_{bend}$ and $C$ represent the normalization,
spectral index above the bend, bending frequency, and Poisson noise level, respectively.
The values of $\alpha$ and $\log(f_{bend})$ are $3.1\pm0.9$ and $-3.5\pm0.3$, respectively.
To check the parameters, we also employ a maximum likelihood method (proposed by \citet{Stella1994,Israel1996,Vaughan2010,Barret2012,Guidorzi2016}) to derive the values of power spectral.
And the parameters of $\alpha$ and $\log(f_{bend})$ are $3.3\pm0.5$ and $-3.3\pm0.2$,  respectively.
The parameters are well in agreement to that found with the $\chi^2$ minimum technique in our work.
Then we employ the method provided in \citet{Emmanoulopoulos2013} to obtain the artificial light-curves,
and evaluate the confidence-curves shown in the lower right panel of Fig.~\ref{powers} (left image).
The green dashed-dotted and blue dashed lines represent the $5\sigma$ and $4\sigma$ confidence levels, respectively.
The confidence level is estimated at $\sim~5.5\sigma$. And Mrk 766 has been monitored 9 times for over $\sim30$ ks with \emph{XMM-Newton}
(in fact the total exposure $\sim0.7$ Ms or $\sim10$ segments of similar length having QPO). While the power peak is independent of frequency bins within its FWHM.
Accounting the number of trials, the confidence level of the QPO is $5.1\sigma$ (99.999965\%).
We also searched for the QPO signal in other observations but found nothing.
This result may reveal that the QPO in NLS1s is a transient phenomenon, consistent with \citet{Gierlinski2008} and \citet{Pan2016}.
Furthermore, the periodic signal in EPIC light-curve is confirmed with the results of 2RGS light-curve at $1.55\times10^{-4}$ Hz,
which is plotted in the right image of Fig.~\ref{powers}.

With the tool \emph{efold} provided in HEASOFT\footnote{https://heasarc.gsfc.nasa.gov/docs/xanadu/xronos/examples/efold.html},
we fold the Segment I of the EPIC light-curve with the period cycle of 6451.6 s, and show it in the lower panel of Fig.~\ref{lc}.
The errors are calculated from {the standard deviation (68.3\%)} of the mean values of each phase bin.
For clarity two cycles are plotted. We fit the folded light-curve with a constant-rate, and derive the reduced $\chi^{2}_{red}=165.6/49$.
The mean count rate is $\sim 25.66~\rm counts~s^{-1}$, and it is shown as red dashed-dotted line in the lower panel of Fig.~\ref{lc}.
From it, we can see that the amplitude of X-ray flux varies with phase clearly.

\subsection{The time-averaged spectra analysis}\label{subsec:sed}
\begin{figure*}
\centering
\includegraphics[angle=0,scale=0.302]{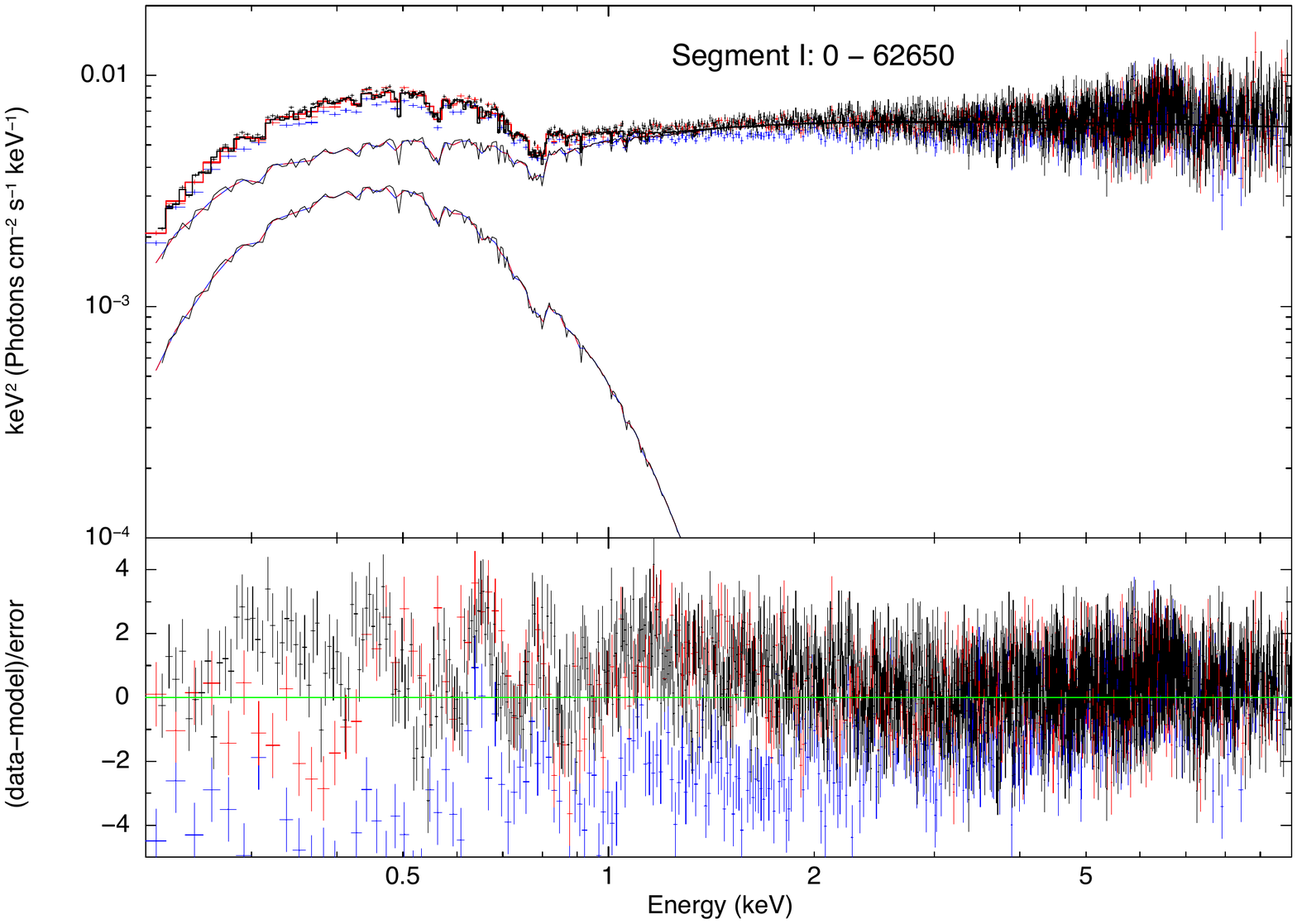}
\includegraphics[angle=0,scale=0.302]{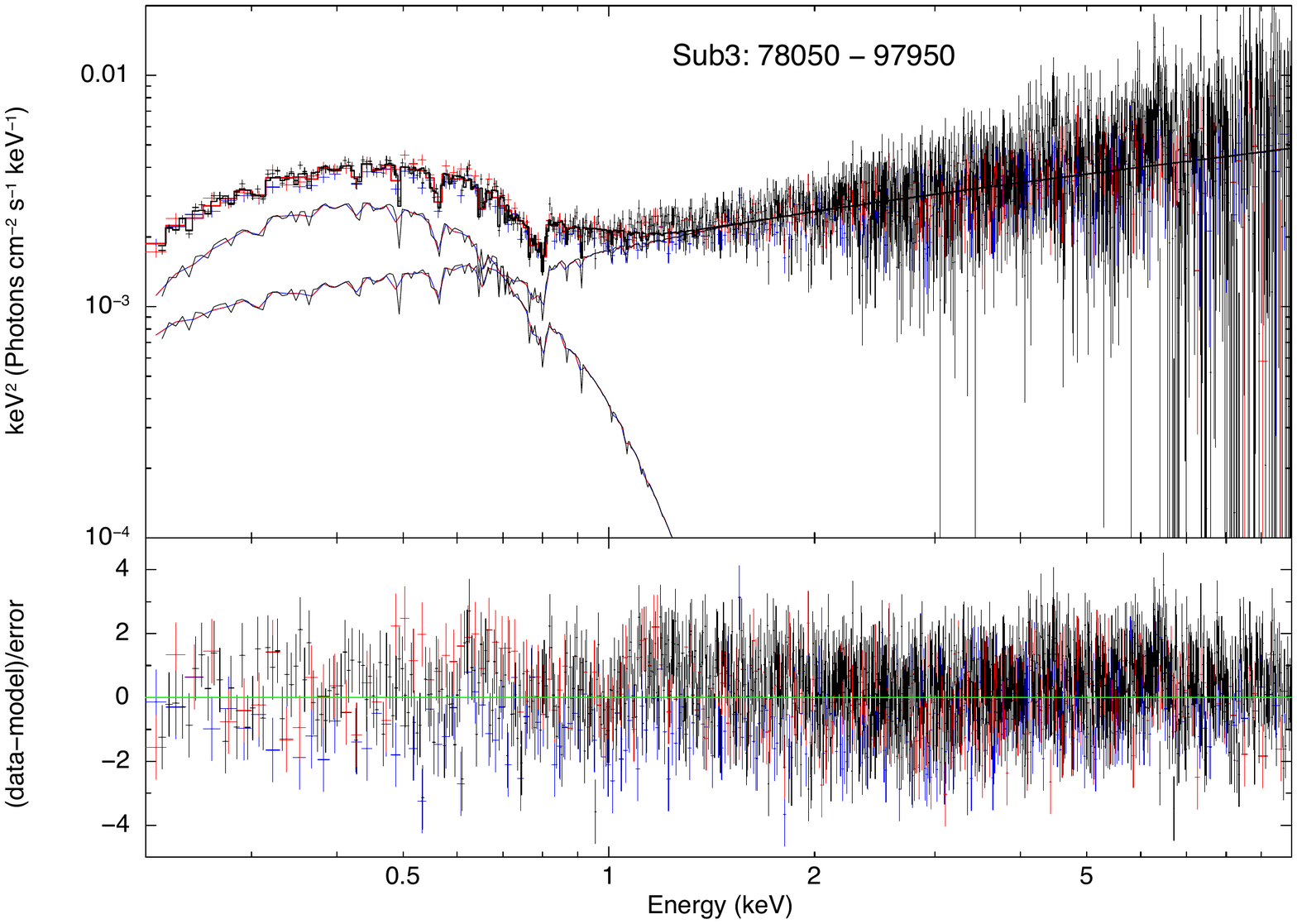}
\includegraphics[angle=0,scale=0.302]{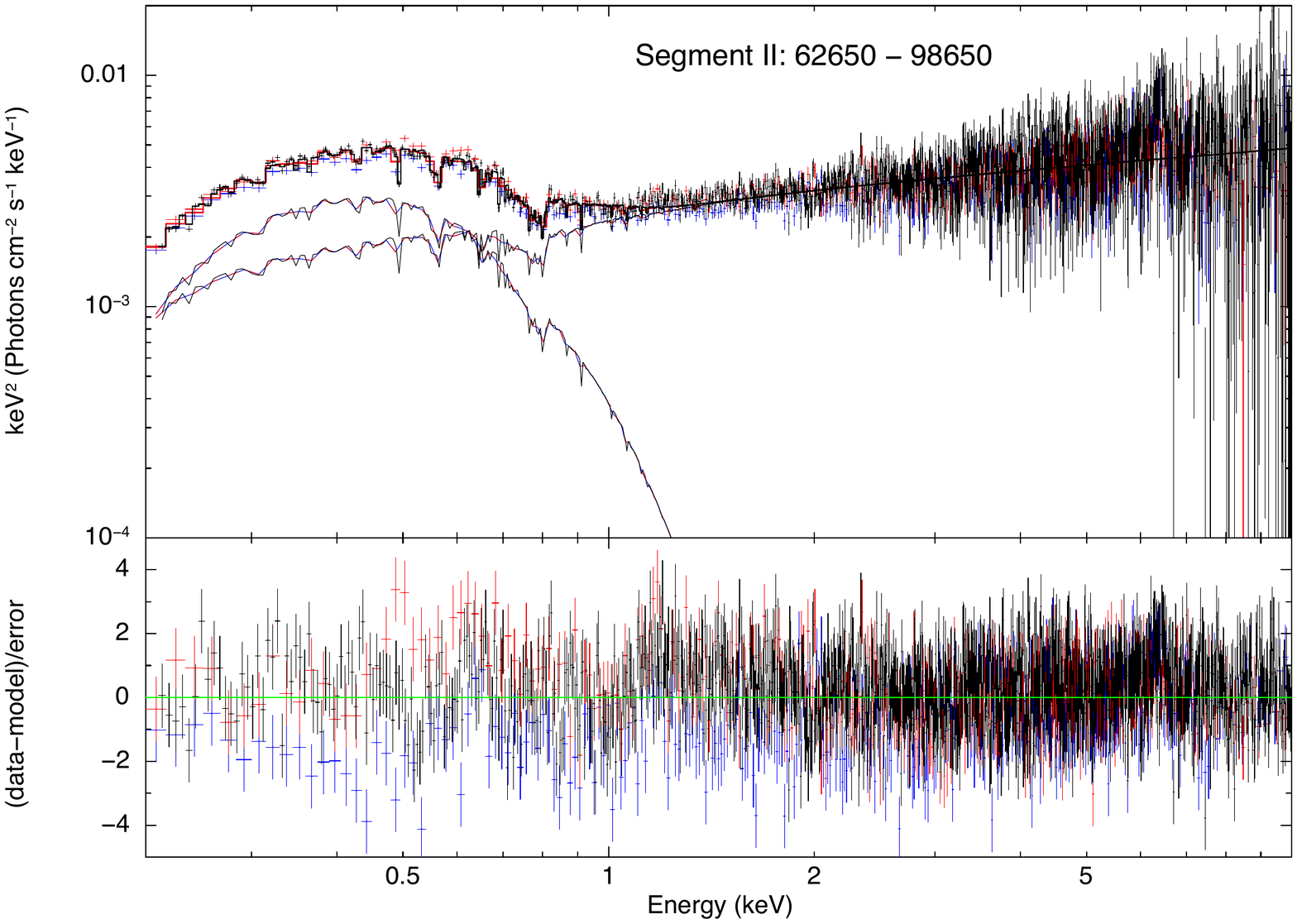}
\includegraphics[angle=0,scale=0.302]{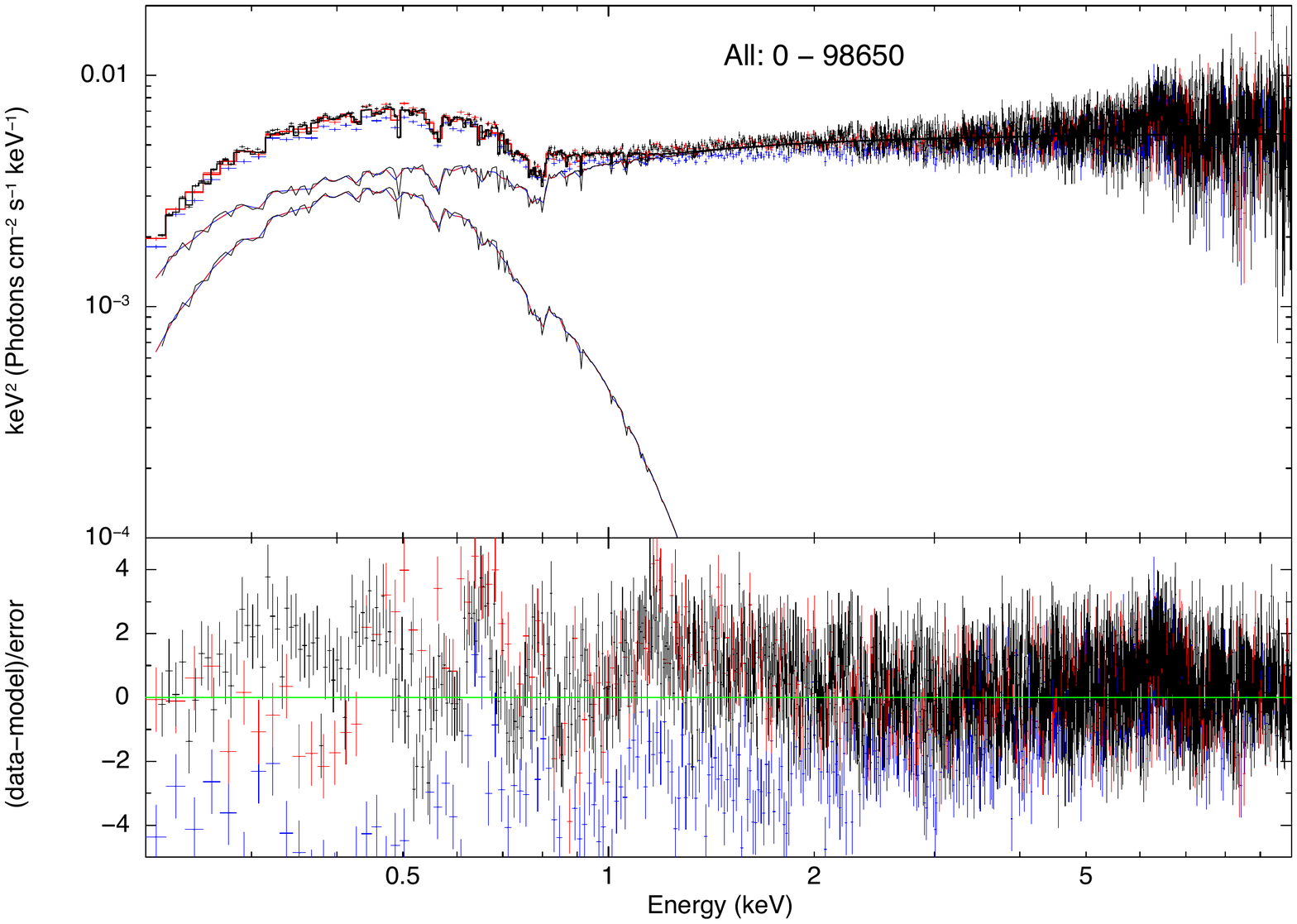}
\caption{The EPIC spectra for four periods are shown in black, red and blue for PN, MOS1 and MOS2, respectively.
              The best-fitting model of $TBabs \times zxipcf \times (zbbody + zpowerlw)$ and the residuals ([data - model]/error) are shown with lines.
              The spectra for Segment I, Sub3, Segment II and the whole period are shown in the upper left, upper right, lower left and lower right images, respectively.}
\label{spec}
\end{figure*}

The spectral analysis is performed using XSPEC \citep[v.~12.9.0n, ][]{Arnaud1996}.
We fit the spectra derived from three EPIC Cameras simultaneously with several models of \emph{TBabs $\times$ zxipcf $\times$ (zbbody + zpowerlw)} \citep{Boller2001} in energy band of 0.2-10 keV.
In the model, \emph{zpowerlw} is a variant of simple power law corrected by redshift of target, which represents the continuum spectrum.
\emph{TBabs} is the Tuebingen-Boulder ISM absorption model representing the Galactic absorption for Mrk 766.
The EPIC spectra indicate clearly the presence of emission of a strong soft excess component below 1 keV.
Then, we employ \emph{zbbody} (a blackbody spectrum with an additional redshift parameter) to fit the strong soft excess,
and the blackbody temperature ($kT_{BB}$) is estimated at $\sim107$ eV (listed in Tab. \ref{Paras}),
which is consistent with the observed temperature of soft excess emission of NLS1s  \citep{Czerny2003,Gierlinski2004}.
Furthermore its emission contains a majority of flux between $0.2-1.0$ keV.
A strong warm absorber is detected at $\sim 1$ keV, we then use an ionized absorber model (\emph{zxipcf}, a model of absorption by partially ionized material) to fit the absorption feature.
All the fitting results for the analysis are acceptable, and the parameters of best-fitting are listed in Tab. \ref{Paras}.
In fitting model to data, we employ $\chi^{2}$ statistic with the errors quoting at 90\% confidence limit.
The four EPIC spectra for all period: $0-98650$ s,  Sub I: $0-62650$ s (with QPO; the very high state),
Segment II: $62650-98650$ s (without QPO), and Sub3: $78050-97950$ s (the lowest flux state) are selected in this analysis.
The best-fitting model and the residuals are shown in the Fig.~\ref{spec}.

\section{SUMMARY AND DISCUSSION}
\label{sec:summary}
In this work, we carry out a systematical analysis of \emph{XMM-Newton} observations of NLS1 Mrk 766 and
detect a QPO signal with a period cycle of $\sim$ 6450 s ($1.55\times10^{-4}$ Hz) at a significance of $>5.1\sigma$
in only part of Segment I (0$-$62650 s) of the observation on 2005 May 31.
And the periodic signal is confirmed subsequently in the data of 2RGS. While in the second part of the X-ray light-curve, no signal is detected at all.
If we use the whole light-curve to analyze, the significance becomes much lower, similar to that found previously in other events \citep[e.g., ][]{Remillard2006,Pan2016,Gonzalez-Martin2012}.
Together with the lack of detection of QPO signals for Mrk 766 in other observations, we suggest that the QPO in NLS1 is likely a transient phenomenon.
In previous works, \citet{Boller2001} reported a possible QPO signal on 2000 May 20 with $2.4\times10^{-4}$ Hz.
The frequency ratio of these two QPO signals, if both valid, is $\sim 3:2$, which would also be the first time to get
such a ratio in X-ray emission of NLS1s. We also analyze the energy spectra derived from EPIC data
and the best-fitting results are listed in Tab.\ref{Paras}.
The ratio of the two periodic signal and the properties of energy spectra are similar to the behaviours of X-ray BHBs.
And the $f_{\rm QPO}$ and $M_{\rm BH}$ of Mrk 766 are consistent with the correlation reported in \citet{Remillard2006}, \citet{Kluzniak2002}, \citet{Zhou2015} and \citet{Pan2016}.

It is widely believed that the QPO signals can be produced by instabilities in the inner accretion disk, or pulsating accretion when it is close to the Eddington limit,
or X-ray hot spots orbiting the BH or disk precession according to Bardeen-Petterson effect
\citep{Remillard2006,Li2004,Sunyaev1973,Guilbert1983,Bardeen1975,Mukhopadhyay2003,Gangopadhyay2012}.
Specifically, in BHB systems, pairs of QPOs have also been detected with frequency ratios of nearly 3:2 \citep{McClintock2006,Strohmayer2001a,Strohmayer2001b,Abramowicz2001}.

\begin{figure*}
\centering
\includegraphics[scale=0.4]{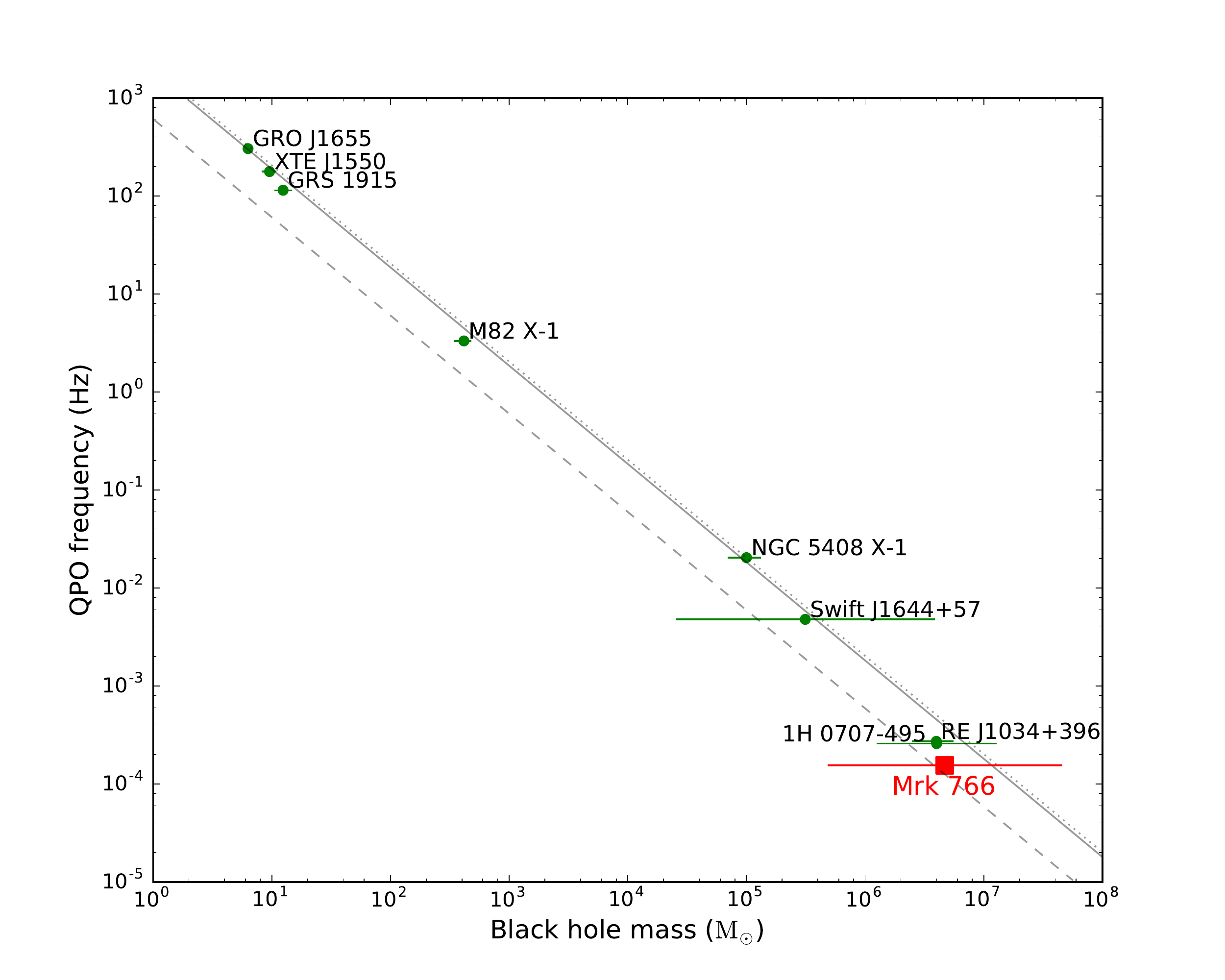}
\caption{The correlation between BH masses and QPO frequencies. The events reported in previous works are shown with green points
              and the new QPO signal detected in Mrk 766 is plotted with a red square. The three lines represent the relations suggested in \citet{Remillard2006} and \citet{Kluzniak2002}.
              Please also see \citet{Zhou2010,Zhou2015} and \citet{Pan2016} for more information.}
\label{fm}
\end{figure*}

The QPO frequencies in RE J1034$+$396 \citep{Gierlinski2008} and 1H 0707$-$495 \citep{Pan2016,Zhang2017}
have been argued to be the High Frequency QPOs \citep[HFQPOs;][]{Zhou2010,Zhou2015}.
The one we found in Mrk 766 is at a similar frequency. Moreover, all of these three sources are NLS1s with similar power-spectral shapes,
strong soft excesses between $0.1-1$ keV in their X-ray energy-spectra and high Eddington ratios.
Hence the signal reported in this work may be a HFQPO.
The correlation of $f_{\rm QPO}\ -\ M_{\rm BH}$ \citep{Remillard2006,Kluzniak2002,Zhou2010,Zhou2015,Pan2016} is shown in Fig.~\ref{fm} and
the QPO in Mrk 766 is well consistent with it, where the mass is adopted from \citet{Turner2006}.
Generally the HFQPOs are only detected in very high states with high accretion rates for X-ray BHBs \citep{Remillard2006,Lai2009}.
Interestingly, the QPOs of NLS1s are also detected at their high state.
The origin of HFQPOs is unclear in X-ray BHBs as well as NLS1s.
Our results reported here may provide us more information for understanding of this phenomena.

The energy spectral fit results indicate that the black body temperatures remain to be a constant at $\sim$ 107 eV within a few percent (listed in Tab. \ref{Paras})
during the four time intervals of the X-ray light-curve shown in Fig.~\ref{lc}.
Comparing the best-fit results of Segment I and Segment II (especially Sub3),
the black body components contributing to flux between 0.2-1.0 keV are remarkably different.
In view of the middle panel of Fig.~\ref{lc} and the Fig.~\ref{powersSII} (i.e., the signal disappeared in Segment II),
we suggest that the presence/absence of the signal are related to the change of the physical process taking place at Mrk 766 rather than the ratio of Signal to Noise.
The similar scenario also detected in galactic X-ray BHBs \citep[e.g., GRO J1655$-$40;][]{McClintock2006,Remillard2006}.
Which may provide an evidence that AGNs are scaled-up versions of Galactic BHBs.

\section*{Acknowledgments}
We thank the anonymous referees for useful and constructive comments.
This work was supported in part by 973 Program of China (No. 2013CB837000),
by NSFC under grants 11525313 (the National Natural Fund for Distinguished Young Scholars)
and 11433009, and by the Key Laboratory of Astroparticle Physics of Yunnan Province (No. 2016DG006).


\clearpage
\begin{table*}
 \centering
 \caption{The spectral parameters of our best-fitting to the data.}
\begin{center}
\scalebox{0.9}{%
\begin{tabular}{c|c|c|c|c|c}
\hline\hline
\multirow{2}{*}{Model component} & \multirow{2}{*}{Parameters}& Segment I & Segment II & Sub3 & Average Spectrum \\ [0.01cm]
 \cline{3-6}
                                                      & & $0 - 62650$ s & $62650 - 98650$ s &$78050 - 97950$ s &$0 - 98650$ s \\  [0.01cm]
\hline\hline
TBabs                               & $N_{H}\ (10^{20}\ \rm cm^{-2})$  & $1.97\pm0.06$  &$0.88\pm0.09$&$0.41\pm0.13$&$1.65\pm0.05$ \\
\hline
\multirow{3}{*}{zxipcf}       & $N_{H}\ (10^{20}\ \rm cm^{-2})$  & $32.78\pm4.88$ &$13.92\pm3.39$&$13.83\pm7.35$ &$22.95\pm2.37$  \\ \cline{2-6}
                                         & $\log\xi\ \rm (erg\ cm\ s^{-1})$ & $0.83\pm0.03$  &$0.81\pm0.06$ & $0.86\pm0.07$ & $0.82\pm0.03$ \\ \cline{2-6}
                                         & C$_f$                                     & $0.62\pm0.04$   &$0.88\pm0.13$& $0.95\pm0.29$ & $0.71\pm0.04$ \\ \cline{2-6}
\hline
\multirow{2}{*}{zbbody}    & $kT_{\rm BB}\ \rm (eV)$               & $107.01\pm1.38$&$106.23\pm1.37$& $108.17\pm1.71$&$106.57\pm1.0$\\ \cline{2-6}
                                         & $Norm\ (\times10^{-4})$                  & $1.25\pm0.03$ &$0.96\pm0.02$& $0.86\pm0.03$ & $1.18\pm0.02$  \\
\hline
\multirow{2}{*}{zpowerlw} & $\Gamma$                             & $2.06\pm0.01$ &$1.76\pm0.01$ & $1.63\pm0.02$& $1.98\pm0.01$   \\ \cline{2-6}
                                         & $Norm_{\rm zpl}\ (\times10^{-3})$         & $7.06\pm0.06$ &$2.83\pm0.04$& $2.11\pm0.04$ & $5.44\pm0.04$   \\
\hline\hline
Reduced $\chi^2$ / $\nu$  & |                                                 & 1.7 / 2532           & 1.2 / 2241& 1.2 / 1871 & 1.9 / 2750  \\
\hline\hline
\end{tabular}}\\
\end{center}
{{\bf Note:} The spectral parameters obtained from the fitting of the time-averaged spectrum and the three time-resolved spectra.
$Norm_{\rm zpl}$ is in units of photons $\rm cm^{-2}~s^{-1}~keV^{-1}$ at 1 keV. The redshift in fitting is fixed to be 0.0127.}
\label{Paras}
\end{table*}
\end{document}